\documentclass[aps,prl,reprint,superscriptaddress,
longbibliography,floatfix]{revtex4-2}

\usepackage[T1]{fontenc}
\usepackage[utf8]{inputenc}
\usepackage{lmodern}
\usepackage{verbatim}

\usepackage{amsmath,amssymb,bm}
\usepackage{physics}
\usepackage{braket}
\usepackage{siunitx}

\usepackage{graphicx}
\usepackage{booktabs}
\usepackage{dcolumn}
\usepackage{xcolor}
\usepackage{soul}
\sethlcolor{yellow}
\usepackage[normalem]{ulem}

\graphicspath{{Figures/workversion/}}
\usepackage{hyperref}
\hypersetup{
  colorlinks=true,
  linkcolor=blue,
  citecolor=blue,
  urlcolor=blue
}

\begin{document}

\title{Direct generation of N00N states with layered semiconductors}

\author{Josip Bajo}
\email{josip.bajo@univie.ac.at}
\affiliation{Vienna Doctoral School in Physics, University of Vienna, Faculty of Physics, 1090 Vienna, Austria}
\affiliation{Vienna Center for Quantum Science and Technology, University of Vienna, Faculty of Physics, 1090 Vienna, Austria}
\author{Benjamin Braun}
\affiliation{Vienna Doctoral School in Physics, University of Vienna, Faculty of Physics, 1090 Vienna, Austria}
\affiliation{Vienna Center for Quantum Science and Technology, University of Vienna, Faculty of Physics, 1090 Vienna, Austria}
\author{Philipp K. Jenke}
\affiliation{Vienna Doctoral School in Physics, University of Vienna, Faculty of Physics, 1090 Vienna, Austria}
\affiliation{Vienna Center for Quantum Science and Technology, University of Vienna, Faculty of Physics, 1090 Vienna, Austria}
\author{Chiara Trovatello}
\affiliation{Department of Mechanical Engineering, Columbia University, New York, NY 10027, USA}
\affiliation{Dipartimento di Fisica, Politecnico di Milano, Piazza L. da Vinci 32, I-20133 Milano, Italy}
\author{Philip Walther}
\affiliation{Vienna Center for Quantum Science and Technology, University of Vienna, Faculty of Physics, 1090 Vienna, Austria}
\affiliation{Christian Doppler Laboratory for Photonic Quantum Computer \& Research Network Quantum Aspects of Space Time (TURIS), University of Vienna, Faculty of Physics, 1090 Vienna, Austria}
\author{Lee A. Rozema}
\affiliation{Vienna Center for Quantum Science and Technology, University of Vienna, Faculty of Physics, 1090 Vienna, Austria}


 \begin{abstract}

\noindent Photonic N00N states enable phase sensing beyond the classical limit, but their generation conventionally relies on interferometric manipulation of photons using bulk optical components. Here we demonstrate the direct generation of a family of tunable two-photon polarization N00N states from a subwavelength-thick 3R-MoS$_2$ flake. For propagation along the optic axis, the in-plane crystal symmetry allows all symmetry permitted spontaneous parametric down-conversion pathway to interfere coherently, while the large second-order nonlinearity provides sufficient brightness at submicrometer thickness without phase matching. The pump polarization sets the basis of the N00N state or equivalently a two-photon Holland-Burnett state without altering its structure or the generation efficiency. The biphotons are highly indistinguishable, reach high fidelity with the target state across a wide range of pump polarizations, and show two-photon interference fringes with visibilities above the threshold for the postselected phase supersensitivity. Our theoretical analysis further indicates that, in the ideal single-mode limit, the same interaction would yield higher-order Holland-Burnett states. By eliminating interferometric stages, our results establish a compact, pump-programmable source for quantum-enhanced polarization interferometry.

\end{abstract}


\maketitle

\noindent N00N states are key resources for quantum-enhanced interferometry: their $N$-photon superposition exhibits $N$-fold phase super-resolution and Heisenberg-limited phase sensitivity~\cite{Dowling2008,Lee2002Rosetta}. They are conventionally prepared by interfering indistinguishable photons, typically from spontaneous parametric down-conversion (SPDC), on beam splitters, often with postselection or heralding~\cite{mitchell2004super,walther2004broglie,edamatsu2002measurement,bogdanov2004qutrit}, 
including in integrated waveguide circuits~\cite{matthews2009manipulation,matthews2011heralding}. 
Such path-entangled states can be directly generated within single nonlinear devices, such as coupled waveguides~\cite{setzpfandt2016tunable,kruse2015dualpath,solntsev2014generation} or structured crystals with shaped pumps~\cite{di2022direct}.
These approaches, however, require additional interferometric stages, engineered pump fields and nonlinear structures, and all have their basis fixed by the phase-matching geometry.
Such considerations, introduce complexity and increase the device footprint, particularly as the photon number grows~\cite{lee2002linear,cable2007efficient}.

Working in a subwavelength nonlinear regime offers a route around these constraints. 
When longitudinal phase matching is relaxed, SPDC pathways that are suppressed in bulk crystals can interfere coherently~\cite{Weissflog2024,rBN2025SPDC,Trovatello2025,braun2026scalingnativeentanglementgeneration}, leaving the emitted quantum state governed directly by the pump field and the material's $\chi^{(2)}$ tensor.
Such relaxed phase matching has been studied theoretically and experimentally in monolayer MoS$_2$~\cite{Saleh2018MoS2SPDC,Ciattoni2018,Trovatello2021}, ultrathin nonlinear films~\cite{Marini2018Constraints,Okoth2019ThinSPDC,SantiagoCruz2021ThinFilmSPDC}, and dielectric metasurfaces~\cite{SantiagoCruz2021Meta}. In an ultrathin GaP film, the pump polarization was shown to tune between entangled and separable states~\cite{sultanov2022flat}, and these ideas were extended to z-cut thin-film lithium niobate~\cite{Shinde2026ThinFilmLN}, atomically layered semiconductors~\cite{Weissflog2024,Trovatello2025}, and atomically thin crystalline metal films~\cite{jenke2026few,abdullah2026plasmon}. Quantum-light generation has now been demonstrated in 3R-MoS$_2$~\cite{Weissflog2024,Trovatello2025,Fan2025MoS2MetasurfaceSPDC,braun2026scalingnativeentanglementgeneration}, WS$_2$~\cite{Feng2024WS2SPDC}, NbOX$_2$ ferroelectrics ($X=\mathrm{Cl,Br,I}$)~\cite{Guo2023NbOCl2,Guo2024NbOCl2Ent,kallioniemi2025van,wang2026nonlinear,prasannan2026entangled}, rhombohedral boron nitride~\cite{rBN2025SPDC,lyu2025tunable}, and GaSe~\cite{lu2025counter,ma2026twin}. These materials are also being integrated into photonic structures~\cite{liu2019van}, including NbOI$_2$ and 3R-MoS$_2$ flakes generating photon pairs directly in single-mode fiber~\cite{joshi2026inlinefiberintegratedphotonpairgeneration,laudert2026demonstrationmonolithicfullytelecomfibercompatible}.

Among these materials, rhombohedrally stacked transition metal dichalcogenides such as 3R-MoS$_2$ are particularly attractive to access this regime: their noncentrosymmetric $C_{3v}$ stacking yields a strong nonlinearity ($\chi^{(2)}\sim\SI{100}{pm/V}$) that builds up coherently over many layers~\cite{shi20173r,xu2022towards}, combined with ease of fabrication~\cite{xu2022towards,Trovatello2025} and compatibility with silicon processing~\cite{kwon2024200,liu2025homoepitaxial}.
However, most demonstrations of SPDC in ultra-thin media separate collinear photon pairs probabilistically on a 50:50 beam splitter to postselect a polarization Bell state~\cite{Weissflog2024,braun2026scalingnativeentanglementgeneration}. This discards at least 50\% of the pairs and leaves a fundamental question open: what two-photon state does the crystal naturally emit into a single spatial mode? Answering this is essential for integrating van der Waals sources directly into single-mode photonic architectures.

\begin{figure}[t]
\centering
\includegraphics[width=\columnwidth]{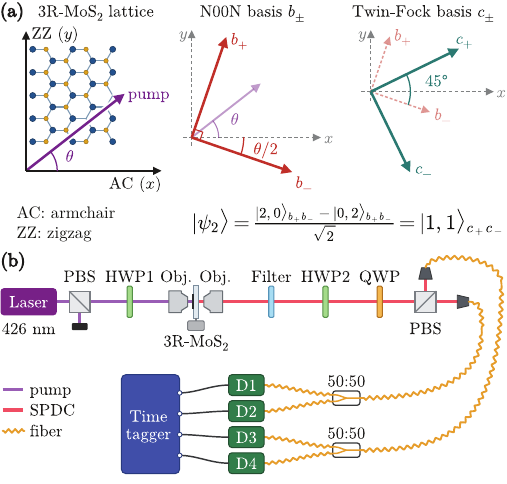}
\caption{\textbf{Direct N00N-state generation in a rhombohedral layered semiconductor.}
(a) 3R-MoS$_2$ lattice and a linearly polarized pump at angle $\theta$. In the pump-selected basis $b_\pm(\theta)$, the source emits $(\ket{2,0}-\ket{0,2})_{b_+,b_-}/\sqrt{2}$; in the complementary basis $c_\pm(\theta)$, rotated by $45^\circ$, the same state is the twin-Fock state (also known as the Holland-Burnett state)  $\ket{1,1}_{c_+,c_-}$.
(b) Setup. A continuous-wave 426-nm pump is focused onto the flake; the filtered emission is analyzed with motorized wave plates and a polarizing beam splitter (PBS). A 50:50 beam splitter (BS) in each PBS output, with four avalanche photodiodes (APDs), provides pseudo-photon-number resolution. HWP (QWP), half- (quarter-) wave plate. Details in the Supplemental Material (SM).}
\label{fig:setup}
\end{figure}

Here, we show that a subwavelength-thick 3R-MoS$_2$ flake directly generates a polarization N00N state without external interferometers. 
Through the interplay of the $C_{3v}$ symmetry and relaxed phase matching, collinear SPDC produces
\begin{equation}
\ket{\Psi_2(\theta)}
=
\frac{1}{\sqrt{2}}
\left[
\ket{2,0}_{b_+(\theta),b_-(\theta)}
-
\ket{0,2}_{b_+(\theta),b_-(\theta)}
\right]
\label{eq:intro_source_noon}
\end{equation}
for any linear pump polarization,
where $b_\pm(\theta)$ are orthogonal polarization modes set by the pump polarization angle $\theta$ relative to the crystal $x$ axis. The pump thus controls the basis in which the N00N state is encoded while preserving its structure [Fig.~\ref{fig:setup}(a)]. 

We use polarization-resolved tomography and phase-sensitive interference to confirm this prediction.
We further show theoretically that the same interaction yields higher-order Holland-Burnett states~\cite{holland1993interferometric,Datta2011HB}.\\

To show how a single crystal emits a N00N state for every pump polarization, consider the symmetry of 3R-MoS$_2$ ($C_{3v}$ point group), whose nonzero in-plane $\chi^{(2)}$ components satisfy

\begin{equation}
\chi^{(2)}_{xxx}
=
-\chi^{(2)}_{xyy}
=
-\chi^{(2)}_{yxy}
=
-\chi^{(2)}_{yyx}
\equiv\chi.
\label{eq:c3v_tensor}
\end{equation}
The first index denotes the pump polarization and the other two the generated photons; $x$ ($y$) lies along the armchair (zigzag) direction, the principal in-plane crystal axes [Fig.~\ref{fig:setup}(a)]. An $x$-polarized pump thus drives $\ket{xx}-\ket{yy}$, whereas a $y$-polarized pump drives $-(\ket{xy}+\ket{yx})$. In a bulk crystal, phase matching would generally select only some of these pathways, and in many entanglement generation schemes group-velocity mismatch must be compensated to make the photons indistinguishable~\cite{kwiat1995new}. 
In a subwavelength flake, all pathways contribute coherently and can interfere. 
For a pump $\mathbf{e}_p(\theta)=\cos\theta\,\mathbf{x}+\sin\theta\,\mathbf{y}$, the pair-creation Hamiltonian is
\begin{equation}
\begin{aligned}
\hat H_I^{(+)}(\theta)
=
\eta\Big\{
&\cos\theta
\left(
\hat a_{s,x}^{\dagger}\hat a_{i,x}^{\dagger}
-
\hat a_{s,y}^{\dagger}\hat a_{i,y}^{\dagger}
\right)
\\
&-
\sin\theta
\left(
\hat a_{s,x}^{\dagger}\hat a_{i,y}^{\dagger}
+
\hat a_{s,y}^{\dagger}\hat a_{i,x}^{\dagger}
\right)
\Big\},
\end{aligned}
\label{eq:signal_idler_hamiltonian}
\end{equation}
where $\eta\propto\chi E_p$ is the pair-creation coupling, including any convention-dependent phase, and $E_p$ is the classical pump amplitude.
In the degenerate, collinear configuration, we assume that both photons occupy the same spatiotemporal mode and differ only in polarization. The signal and idler labels then become redundant, $\hat a_{s,\mu}^{\dagger},\hat a_{i,\mu}^{\dagger}\rightarrow\hat a_{\mu}^{\dagger}$ ($\mu=x,y$), and Eq.~\eqref{eq:signal_idler_hamiltonian} reduces to
\begin{equation}
\hat H_I^{(+)}(\theta)
=
\eta
\left\{
\cos\theta
\left[
(\hat a_x^\dagger)^2-(\hat a_y^\dagger)^2
\right]
-
2\sin\theta\,
\hat a_x^\dagger\hat a_y^\dagger
\right\},
\label{eq:fixed_basis_hamiltonian}
\end{equation}
where the $xy$ and $yx$ amplitudes now add coherently. This generates
\begin{equation}
\ket{\Psi_2(\theta)}=
\frac{\cos\theta}{\sqrt{2}}
(\ket{2,0}_{x,y}-\ket{0,2}_{x,y})
-\sin\theta\ket{1,1}_{x,y},
\label{eq:fixed_basis_state}
\end{equation}
which for $\theta=0$ is a N00N state in the crystal basis. For arbitrary $\theta$, we introduce the pump-selected modes
\begin{equation}
\begin{aligned}
\hat a^\dagger_{b_+(\theta)}
&=
\cos\frac{\theta}{2}\,\hat a_x^\dagger
-
\sin\frac{\theta}{2}\,\hat a_y^\dagger,
\\
\hat a^\dagger_{b_-(\theta)}
&=
\sin\frac{\theta}{2}\,\hat a_x^\dagger
+
\cos\frac{\theta}{2}\,\hat a_y^\dagger,
\end{aligned}
\label{eq:pump_selected_basis}
\end{equation}
in which
\begin{equation}
\hat H_I^{(+)}(\theta)
=\eta\left[\left(\hat a^\dagger_{b_+(\theta)}\right)^2-\left(\hat a^\dagger_{b_-(\theta)}\right)^2\right],
\label{eq:source_noon_state}
\end{equation}
which generates exactly the state of Eq.~\eqref{eq:intro_source_noon}. The $\chi^{(2)}$ tensor thus fixes the amplitudes and relative phase of the N00N superposition, while the pump only sets its basis: rotating the pump polarization by $\theta$ rotates the N00N basis by $-\theta/2$ [Eq.~\eqref{eq:pump_selected_basis}], leaving the structure of the state unchanged for any linear pump polarization.

These states are related to the twin-Fock states, also known as Holland-Burnett states in the context of metrology~\cite{holland1993interferometric}, through a simple change of basis (Fig.~\ref{fig:setup}(a)). 
In the complementary modes
\begin{equation}
\hat a^\dagger_{c_\pm(\theta)}
=
\frac{
\hat a^\dagger_{b_+(\theta)}
\pm
\hat a^\dagger_{b_-(\theta)}
}{\sqrt{2}},
\label{eq:complementary_basis}
\end{equation}
the Hamiltonian becomes $\hat H_I^{(+)}=2\eta\,\hat a^\dagger_{c_+}\hat a^\dagger_{c_-}$ which generates the two-photon Holland-Burnett state $\ket{1,1}_{c_+(\theta),c_-(\theta)}$. As in Hong-Ou-Mandel interference, a balanced mode transformation connects two-photon Holland-Burnett states and two-photon N00N states~\cite{chen2018polarization}; here, we describe this merely a change of polarization basis. More generally, the 2n-photon Holland-Burnett state $\ket{n,n}_{c}$ reaches Heisenberg scaling of the phase sensitivity while being considerably more robust to loss than N00N states~\cite{holland1993interferometric,Datta2011HB}. Because the Hamiltonian has the two-mode-squeezing form in the $c_\pm$ basis, in the single-mode limit and conditioned on $n$ pairs the source emits
\begin{equation}
\ket{\Psi_{2n}^{c}(\theta)} = \ket{n,n}_{c_{+}(\theta),c_{-}(\theta)},
\label{eq:higher_order_state}
\end{equation}
i.e., a $2n$-photon Holland-Burnett state in the $c_\pm$ basis for any linear pump polarization (derivation in SM). For $n\geq2$, these states contain additional photon-number components and differ from N00N states~\cite{thomas2011integrated}, but remain valuable for quantum metrology, with enhanced phase sensitivity demonstrated up to six photons~\cite{xiang2013optimal,jin2016detection}.

\begin{figure}[!t]
\centering
\includegraphics[width=\columnwidth]{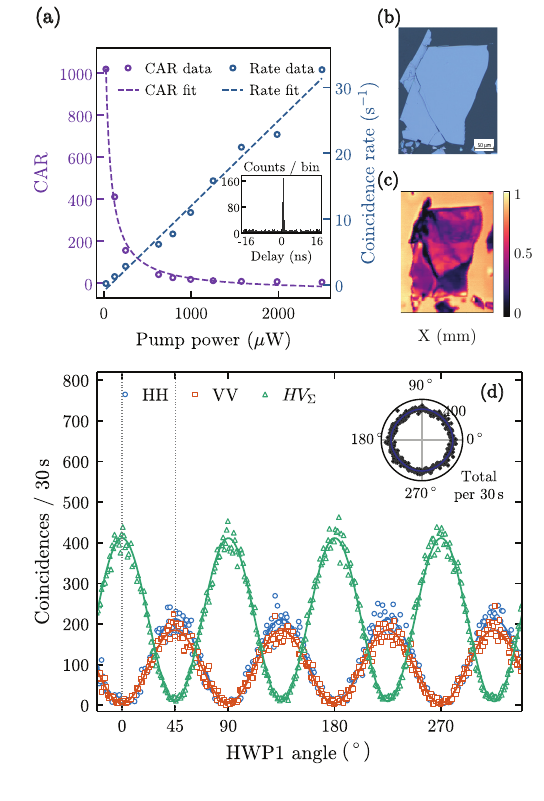}
\caption{\textbf{SPDC Characterization in 3R-MoS$_2$.}
(a) Coincidence rate and coincidence-to-accidental ratio (CAR) as functions of pump power. \textit{Inset}: representative coincidence time-difference histogram with a peak at $\Delta t=0$.
(b) Optical micrograph of the measured 285-nm thick flake.
(c) Spatial emission map, showing localization of the detected signal to the flake.
(d) Coincidence rates in the $HH$, $VV$, and combined $HV_\Sigma$ channels as functions of the pump-HWP1 angle. \textit{Inset}: polarization-summed coincidence response, which remains
approximately constant during pump rotation.}
\label{fig:fig2}
\end{figure}

The single-mode assumption can be tested experimentally, because it constrains the polarization: bosonic exchange symmetry restricts the two photons to the symmetric spin-1 triplet $\{\ket{HH},\ket{\Psi^+},\ket{VV}\}$, with $\ket{\Psi^\pm}=(\ket{HV}\pm\ket{VH})/\sqrt{2}$, forming a biphoton polarization qutrit~\cite{bogdanov2004qutrit}. In the spin picture, each photon is a spin-$1/2$ ($\ket{H},\ket{V}\leftrightarrow m=\pm\tfrac12$), and $\ket{HH},\ket{\Psi^+},\ket{VV}$ are the $m=+1,0,-1$ triplet states. The antisymmetric singlet $\ket{\Psi^-}$ can be populated only if another degree of freedom is also antisymmetric under exchange~\cite{Adamson2008}. Its population $P_{\Psi^-}=\bra{\Psi^-}\rho\ket{\Psi^-}$ therefore diagnoses residual distinguishability, i.e., a breakdown of the single-mode approximation.

\begin{figure*}[!t]
\centering
\includegraphics[width=1.9\columnwidth]{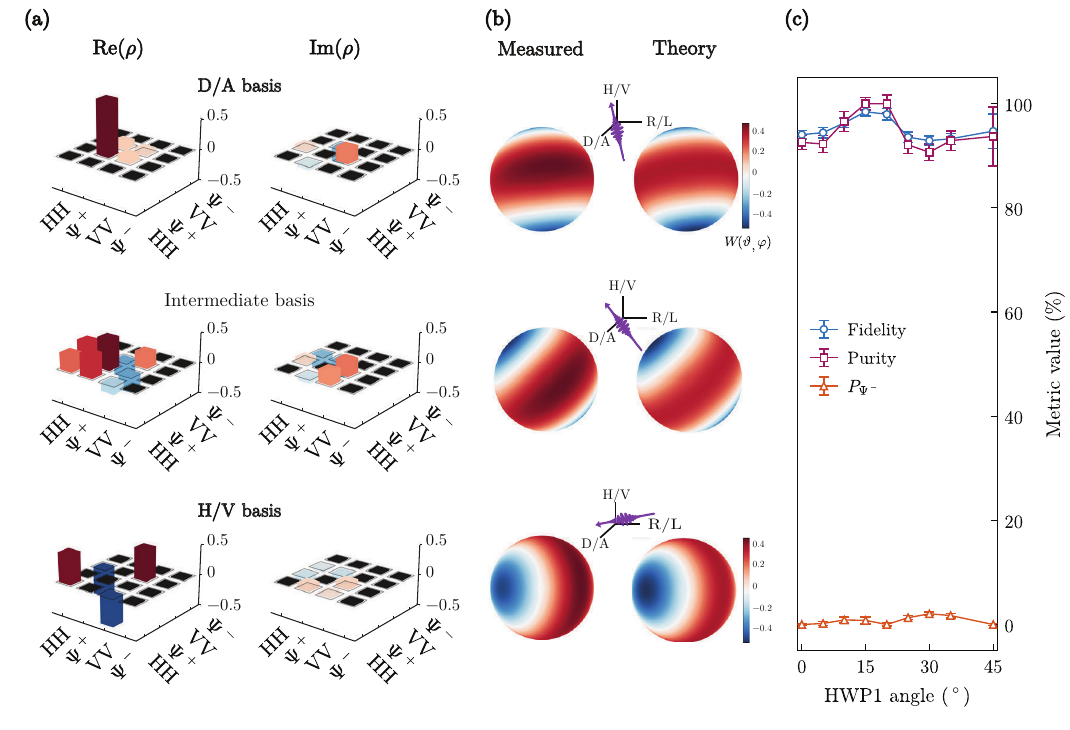}
\caption{\textbf{Tomography of Biphoton States.}
(a) Real and imaginary parts of reconstructed density matrices in the $H/V$ basis, showing pump-dependent populations and coherences.
(b) Measured and theoretical spin-1 Wigner functions on the Poincar\'e sphere; the purple arrow indicates the pump polarization.
For panels (a) and (b), the pump polarization was set to $0^\circ$, $40^\circ$ and $90^\circ$ for the top, middle and bottom rows, respectively.
(c) Target-state fidelity, symmetric-sector purity, and antisymmetric population $P_{\Psi^-}$ versus pump-HWP1 angle.
}
\label{fig:tomography}
 \end{figure*}

We test these predictions with the setup in Fig.~\ref{fig:setup}(b). A continuous-wave 426-nm laser pumps a mechanically exfoliated, $\approx$285-nm-thick 3R-MoS$_2$ flake on fused silica [Fig.~\ref{fig:fig2}(b)]. A spatial scan of the collected signal [Fig.~\ref{fig:fig2}(c)] reproduces the flake features seen in the micrograph; all further measurements are taken near the flake center, where the coincidence-to-accidental ratio (CAR) is highest. At normal incidence, all fields propagate along the optic axis, where 3R-MoS$_2$ is in-plane isotropic, so both down-converted polarizations see the same refractive index~\cite{braun2026scalingnativeentanglementgeneration}; with the subwavelength interaction length, residual out-of-plane effects of the finite focusing and collection NA are minimal. The degenerate pairs, centered at 852~nm, are collected in transmission and spectrally filtered. For initial characterization, the PBS is removed and the emission is split by a 50:50 fiber beam splitter onto two silicon avalanche photodiodes.

We first confirm the SPDC origin of the light: the detection-time-difference histogram shows a pronounced peak at zero delay [inset of Fig.~\ref{fig:fig2}(a)], identifying time-correlated photon pairs. The coincidence rate grows linearly with pump power, while the CAR decreases from $\approx10^3$ at 25\,$\mu$W to $\approx5$ at 2.5\,mW, mainly due to photoluminescence~\cite{Feng2024WS2SPDC}. After accidental subtraction, the detected pair rate is $(10.2\pm0.4)$\,Hz\,mW$^{-1}$ [$\approx0.25$\,Hz\,mW$^{-1}$\,nm$^{-1}$ in the 40-nm band (SM)], comparable to the highest values reported for van der Waals SPDC sources under visible excitation~\cite{rBN2025SPDC,Feng2024WS2SPDC}, while periodic poling has been shown to increase it further~\cite{Trovatello2025,braun2026scalingnativeentanglementgeneration}. 

 \begin{figure}[t]
\centering
\includegraphics[width=\columnwidth]{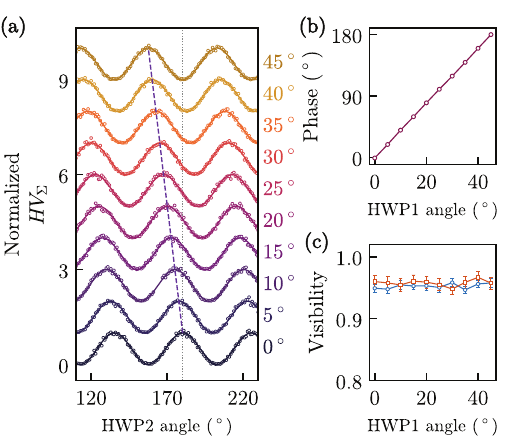}
\caption{\textbf{Phase-sensitive interference.}
(a) Normalized coincidence fringes versus analyzer-HWP2 angle for several pump-HWP1 settings.
(b) Fitted fringe phase versus HWP1 angle; its linear evolution reflects the rotation of the $b_\pm(\theta)$ modes relative to the analyzer.
(c) Fringe visibilities of the cross- (blue) and same-polarization (orange) coincidences.}
\label{fig:fig4}
\end{figure}
For polarization analysis, motorized half- and quarter-wave plates set the measurement basis before a PBS [Fig.~\ref{fig:setup}(b)]. Each PBS output is split by a 50:50 fiber beam splitter onto two detectors, providing pseudo-photon-number resolution: a coincidence between the two detectors of one output projects onto two photons in the same mode, e.g., $\ket{HH}$ or $\ket{VV}$ with the wave plates at $0^\circ$, whereas any of the four cross-output pairs projects onto $\Pi_{HV_\Sigma}=\ket{\Psi^+}\bra{\Psi^+}+\ket{\Psi^-}\bra{\Psi^-}$. A tomographically complete set of wave-plate settings (SM) can be used to also test for hidden distinguishability~\cite{Adamson2007Hidden,Adamson2008}; i.e. population in the antisymmetric subspace. Rotating the pump half-wave plate (HWP1) while recording the $HH$, $VV$, and $HV_\Sigma$ coincidences [Fig.~\ref{fig:fig2}(d)] shows that, as expected, the total SPDC rate is approximately independent of the linear pump polarization [inset]~\cite{Trovatello2021,Weissflog2024}, whereas the emitted polarization changes dramatically. Near $0^\circ$, $HV_\Sigma$ dominates, consistent with the Holland-Burnett state $\ket{1,1}_{H,V}$; near $45^\circ$ ($90^\circ$ pump polarization rotation), $HH$ and $VV$ dominate, consistent with $(\ket{2,0}_{H,V}-\ket{0,2}_{H,V})/\sqrt{2}$; intermediate settings continuously redistribute the rates.

These populations alone, however, cannot distinguish a N00N state from an incoherent mixture of $\ket{HH}$ and $\ket{VV}$. We therefore reconstruct the $3\times3$ symmetric block in the basis $\{\ket{HH},\ket{\Psi^+},\ket{VV}\}$ together with $P_{\Psi^-}$, implementing hidden-differences tomography~\cite{Adamson2007Hidden} (details in SM). Figure~\ref{fig:tomography}(a) shows representative density matrices in the $H/V$ basis. With HWP1 at $0^\circ$ (top), the state is dominated by $\ket{\Psi^+}$, as expected. At $45^\circ$ (bottom), the $HH$ and $VV$ populations are nearly equal, with a negative coherence indicating the N00N state $(\ket{HH}-\ket{VV})/\sqrt{2}$. As argued above, these two states are in fact equivalent: the top state is essentially $\ket{1,1}_{H,V}$, which can be written as $(\ket{2,0}_{D,A}-\ket{0,2}_{D,A})/\sqrt{2}$. At $20^\circ$ (middle row), populations and coherences appear across all three components; the N00N structure is hidden in the $H/V$ representation, but the state is consistent with Eq.~\eqref{eq:intro_source_noon} in the pump-defined $b_\pm(\theta)$ basis.

Figure~\ref{fig:tomography}(c) summarizes ten pump settings. The fidelity with the ideal N00N state in the corresponding $b_\pm(\theta)$ basis (circles) and the symmetric-sector purity (squares) average $(95.00\pm1.94)\%$ and $(94.28\pm3.38)\%$, respectively (standard deviations across settings; error bars from Monte Carlo resampling of the counts). The key test of whether photons from different tensor elements are indistinguishable is the antisymmetric population: ideally zero, its mean is only $(0.69\pm0.78)\%$, confirming that the states lie in the symmetric manifold, i.e., that the photons are highly indistinguishable and the single-mode approximation holds. All tomography used a 3-nm filter to reduce spectral averaging, with CAR values between 35 and 40. We attribute the remaining variations mainly to polarization-dependent coupling into the multimode fiber.

Figure~\ref{fig:tomography}(b) shows the reconstructed states as spin-1 Wigner functions on the Poincar\'e sphere, the SU(2) analog of the conventional phase-space Wigner function~\cite{agarwal1981relation,dowling1994wigner, luis2005quantum,goldberg2021quantum,sanchez2025phase}. This representation has been used for few-photon polarization states~\cite{shalm2009squeezing, rozema2014optimizing,ferretti2024generating} and atomic and solid-state spin systems~\cite{mcconnell2015entanglement, yu2025schrodinger}. While the two-photon N00N state has a vanishing mean Stokes
vector, $\langle\hat{\mathbf{S}}\rangle=0$, its second-order polarization correlations are encoded in the quadrupolar structure of the Wigner function~\cite{bjork2012central, klyshko1997polarization,goldberg2021quantum}, which appears as a positive ring with negative caps on either side~\cite{shalm2009squeezing}. A change of polarization basis, which, for example, maps the N00N state onto a Holland-Burnett state, merely rotates the Wigner function on the sphere. The two representations thus describe the same nonclassical polarization state in different bases~\cite{goldberg2021quantum,ares2024photonic}. The measured Wigner functions rotate rigidly with the pump polarization (purple arrow), showing directly that the N00N basis follows the pump.

Finally, we ask whether this coherence survives a transformation applied \textit{after} generation, by rotating HWP2 while recording the cross-polarized coincidences $HV_\Sigma$ summed over all four $H$-$V$ detector pairs [Fig.~\ref{fig:fig4}(a)]. Rotating HWP1 by $\alpha$ rotates the pump by $2\alpha$ and hence, by Eq.~\eqref{eq:pump_selected_basis}, the N00N basis by $-\alpha$; since a two-photon N00N fringe oscillates at four times the polarization rotation angle, twice as fast as a classical or single-photon fringe~\cite{edamatsu2002measurement,mitchell2004super,Dowling2008}, its phase should shift by $4\alpha$. Indeed, as HWP1 is rotated from $0^\circ$ to $45^\circ$, the fitted fringe phase advances approximately linearly by $\approx180^\circ$ [Fig.~\ref{fig:fig4}(b)]. This reflects the rotation of the N00N basis relative to the fixed analyzer, not a change of the intrinsic relative phase. The fringe visibilities remain above 95\% for all pump settings [Fig.~\ref{fig:fig4}(c)] ,well above the visibility threshold  required for two-photon phase supersensitivity \cite{muller2017quantum}.

Returning to the question posed at the introduction, we have shown that the native nonlinear response of a subwavelength semiconductor with suitable crystal symmetry can itself define a useful entangled state. Relaxed phase matching in 3R-MoS$_2$ enables coherent interference between symmetry-allowed SPDC whose basis is controlled by the pump polarization. In the ideal single-mode limit, the same mechanism generates higher-order Holland--Burnett states. Although pair rates in ultrathin sources remain low, materials with higher damage thresholds, such as
rhombohedral boron nitride~\cite{rBN2025SPDC,lyu2025tunable}, combined with resonant nanostructuring~\cite{liu2019van}
and periodic poling~\cite{Trovatello2025, braun2026scalingnativeentanglementgeneration}, could bring the multiphoton regime within useful quantum technology reach. Our results establish crystal symmetry as a resource for quantum-state engineering without additional interferometric stages.
\\

\begin{acknowledgments}
\noindent \textit{Acknowledgments }
We thank L. K. Shalm for assistance with the Wigner function plotting code way back in 2009.
 This research was funded in whole or in part by the Austrian Science Fund (FWF) [10.55776/COE1] (Quantum Science Austria), [10.55776/F71] (BeyondC) and [10.55776/FG5] (Research Group 5). For open access purposes, the author has applied a CC BY public copyright license to any author accepted manuscript version arising from this submission. P.W. acknowledges financial support from the Air Force Office of Scientific Research under award number FA9550-21-1-0355 (Q-Trust) and FA8655-23-1-7063 (TIQI), as well as the Austrian Federal Ministry of Labour and Economy, the National Foundation for Research, Technology and Development and the Christian Doppler Research Association.  J.B., B.B. and P.K.J. acknowledge support from the University of Vienna via the Vienna Doctoral School. 
 C.T. acknowledges the European Union’s Horizon Europe research and innovation programme under the Marie Skłodowska-Curie PIONEER HORIZON-MSCA-2021-PF-GF grant agreement No 101066108. C.T. also acknowledges the Optica Foundation for supporting this research through the 2024 Optica Foundation Challenge.
\\
\\
\textit{Data availability }— The experimental data supporting the findings of this study are publicly available in the Zenodo repository at \url{https://doi.org/10.5281/zenodo.24309887}.
\end{acknowledgments}

\bibliographystyle{apsrev4-2}
\bibliography{references3jab}

\end{document}


\renewcommand{\thefigure}{S\arabic{figure}}
\renewcommand{\thetable}{S\Roman{table}}

\title{Supplemental Material for:\\
Direct generation of N00N states with layered semiconductors}

\author{Josip Bajo}
\email{josip.bajo@univie.ac.at}
\affiliation{Vienna Doctoral School in Physics, University of Vienna, Faculty of Physics, 1090 Vienna, Austria}
\affiliation{Vienna Center for Quantum Science and Technology, University of Vienna, Faculty of Physics, 1090 Vienna, Austria}

\author{Benjamin Braun}
\affiliation{Vienna Doctoral School in Physics, University of Vienna, Faculty of Physics, 1090 Vienna, Austria}
\affiliation{Vienna Center for Quantum Science and Technology, University of Vienna, Faculty of Physics, 1090 Vienna, Austria}

\author{Philipp K. Jenke}
\affiliation{Vienna Doctoral School in Physics, University of Vienna, Faculty of Physics, 1090 Vienna, Austria}
\affiliation{Vienna Center for Quantum Science and Technology, University of Vienna, Faculty of Physics, 1090 Vienna, Austria}

\author{Chiara Trovatello}
\affiliation{Department of Mechanical Engineering, Columbia University, New York, NY 10027, USA}
\affiliation{Dipartimento di Fisica, Politecnico di Milano, Piazza L. da Vinci 32, I-20133 Milano, Italy}

\author{Philip Walther}
\affiliation{Vienna Center for Quantum Science and Technology, University of Vienna, Faculty of Physics, 1090 Vienna, Austria}
\affiliation{Christian Doppler Laboratory for Photonic Quantum Computer \& Research Network Quantum Aspects of Space Time (TURIS), University of Vienna, Faculty of Physics, 1090 Vienna, Austria}

\author{Lee A. Rozema}
\affiliation{Vienna Center for Quantum Science and Technology, University of Vienna, Faculty of Physics, 1090 Vienna, Austria}

\date{\today}
\maketitle
\tableofcontents
\newpage
\renewcommand{\labelenumi}{\Roman{enumi}.}

\section{Experimental setup and measurement protocol}
\label{sec:setup}
The experiment is based on spontaneous parametric down-conversion (SPDC) in a 3R-MoS$_2$ flake driven by a continuous-wave pump laser at 426~nm (TOPTICA DL Pro). The laser is out coupled through a single-mode fiber and passes through a polarizing beam splitter and a half-wave plate, which prepares a well defined input polarization. Rotation of the input half-wave plate tunes the pump polarization and therefore the source-generated two-photon state.

The pump is focused onto the sample with an achromatic lens (Bernard Halle, $f=33$~mm, clear aperture $4.7$~mm), resulting in an estimated focal spot of approximately $4~\mu$m. The sample is mounted in transmission geometry on a piezoelectric stage (across x, y, z directions), allowing spatial scans across the flake and optimization of the excitation position.

The emitted photons are collected with a $20\times$ objective (OptoSigma PAL-20-NIR-HR-LC00, NA $=0.45$, $f=10$~mm, working distance $20$~mm).  Residual pump light is suppressed with two hardcoated 450~nm long-pass filters (OD6). The SPDC bandwidth is selected with band-pass filters centered near 850~nm, with bandwidths of 40~nm, 10~nm, or 3~nm depending on the measurement.

After spectral filtering, motorized half and quarter-wave plates define the polarization analysis basis. A polarizing beam splitter separates the analyzed polarization components. The polarization analysis in our setup can be described using the standard Jones formalism~\cite{ChekhovaBanzer2021}, where the wave plates define the measurement basis and the polarizing beam splitter separates the two orthogonal polarization components. The photons are coupled into multimode fibers with a 50~$\mu$m core and detected with Excelitas single-photon avalanche detectors. Arrival times are recorded using a multichannel time tagger. This time-tagged record is used to identify coincidences, i.e., detection events on two channels that occur within a chosen coincidence window of approximately 1.054 ns. 

\begin{table}[t]
\caption{\label{tab:components}Key components of the experimental setup.}
\begin{ruledtabular}
\begin{tabular}{p{0.30\columnwidth}p{0.62\columnwidth}}
Component & Specification / role \\
\hline
Pump laser & Continuous-wave laser, 426~nm (TOPTICA DL Pro) \\
Polarization preparation & Polarizing beam splitter and achromatic half-wave plate for 426~nm \\
Focusing optics & Achromatic lens, $f=33$~mm, clear aperture $4.7$~mm, estimated NA $\approx 0.09$ \\
Sample & 3R-MoS$_2$, thickness $\sim 285$~nm, on fused silica, mechanically exfoliated from bulk crystal (HQ Graphene) \\
Collection optics & OptoSigma PAL-20-NIR-HR-LC00, $20\times$, NA $=0.45$, $f=10$~mm, WD $=20$~mm \\
Filtering & Thorlabs FELH450 long-pass filters; FBH850-40, FBH850-10, and FBH852-3 band-pass filters \\
Polarization analysis & Motorized Elliptec stages with achromatic QWP/HWP and polarizing beam splitter \\
Fiber coupling & Multimode fiber coupling, 50~$\mu$m core \\
Single-photon detectors & Excelitas SPCM-NIR and SPCM-AQRH detectors, D1--D4 \\
Time tagging & UQD multichannel time tagger, nominal resolution 156.5~ps \\
\end{tabular}
\end{ruledtabular}
\end{table}

\begin{table}[t]
\caption{\label{tab:angles}Analyzer settings used for polarization tomography.}
\begin{ruledtabular}
\begin{tabular}{ccc}
Setting & HWP angle ($^\circ$) & QWP angle ($^\circ$) \\
\hline
1 & 0    & 0 \\
2 & 22.5 & 0 \\
3 & 0    & 22.5 \\
4 & 22.5 & 22.5 \\
5 & 0    & 45 \\
6 & 22.5 & 45 \\
7 & 45   & 0 \\
\end{tabular}
\end{ruledtabular}
\end{table}

\section{Source calibration, power scaling, and coincidence histograms}
\label{sec:source}

The source was characterized through pump-power scans, spatial scans across the flake (by moving motorized stage), coincidence histograms, and polarization-resolved coincidence measurements. These measurements confirm the nonlinear origin of the signal and define the operating points used for tomography and interference measurements in  Fig.2.

\subsection{Coincidence rates, accidentals, and CAR}
\begin{figure}[t]
\centering
\includegraphics[width=\columnwidth]{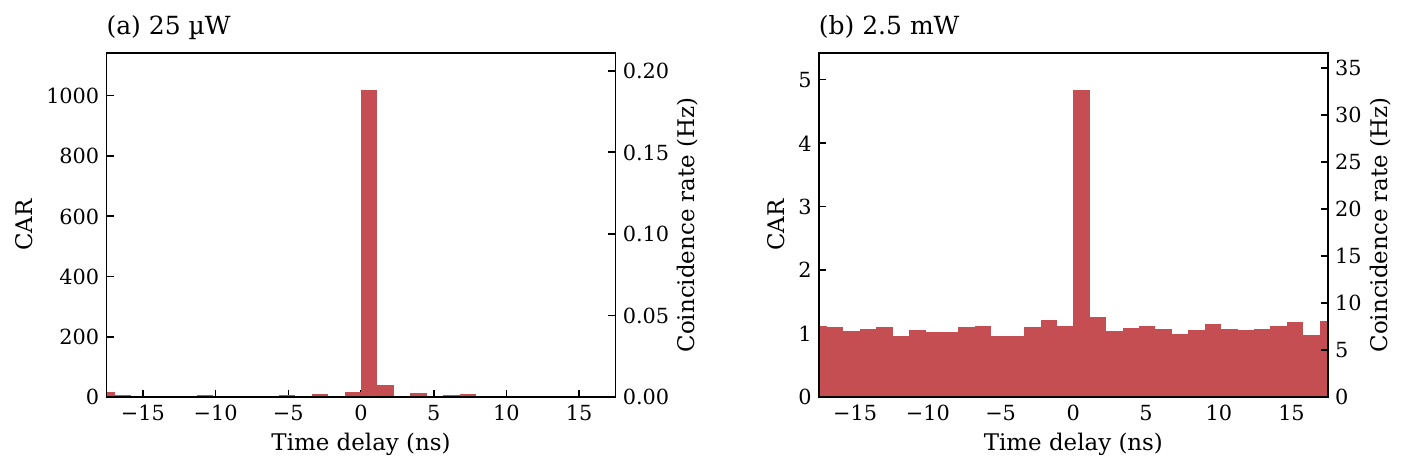}
\caption{\label{fig:car_hists}
Representative coincidence histograms from the highest and lowest value of power series. The coincidence peak is centered at the relative delay for which the two detector channels register paired photons. Counts away from the peak provide an estimate of the accidental background.}
\end{figure}

For each measurement, single count rates $S_i$ and coincidence rates $C_{ij}$ were extracted for the active detector channels. The index $i$ labels one detector channel, while $j$ labels another detector channel. Thus, $S_i$ is the rate of single-photon detections on channel $i$, and $C_{ij}$ is the rate of two-channel events between detectors $i$ and $j$. Accidentals were estimated and subtracted as:
\begin{equation}
A_{ij}=S_iS_j\tau,
\end{equation}
where $A_{ij}$ is the accidental coincidence rate and $\tau$ is the effective coincidence window.
The coincidence-to-accidental ratio is
\begin{equation}
\mathrm{CAR}_{ij}=\frac{C_{ij}}{A_{ij}}.
\end{equation}
Here, CAR is a dimensionless quantity. Representative histograms (Fig.~\ref{fig:car_hists}) show the appearance of a well defined coincidence peak above the accidental background. Accidental coincidences were subtracted from all reported coincidence data, except for the power-dependent measurements shown in Fig.2(a), which are presented without background subtraction.

\subsection{Power scaling}

\begin{figure}[t]
\centering
\includegraphics[width=\columnwidth]{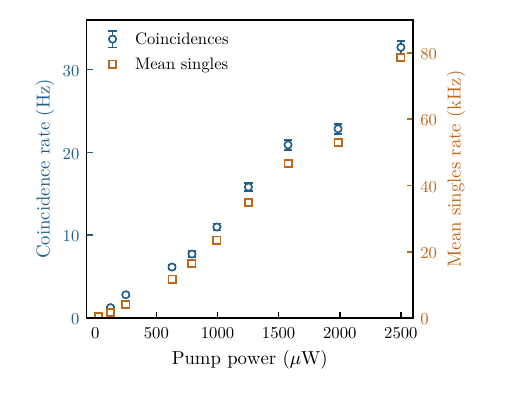}
\caption{\label{fig:singles_cc}
Pump-power dependence of photon detection rates. Measured coincidence and singles rates as functions
of pump power. The linear fit to the accidental-subtracted coincidence rate is used to determine the detected brightness.}
\end{figure}

Pump-power scans were acquired by varying the pump power with a neutral-density filter wheel. In the low-gain SPDC regime, the true-pair generation rate scales linearly with pump power, while accidentals and multipair contributions reduce the CAR at higher powers.  The coincidence count rate was fitted with a linear model, consistent with the expected low-gain SPDC scaling. The CAR was fitted with the  hyperbolic form
\[
\mathrm{CAR}(P)=\frac{\alpha}{P+\beta}+\gamma ,
\]
where \(P\) is the pump power and \(\alpha\), \(\beta\), and \(\gamma\) are free fit parameters. The coincidence rates and CAR values shown in Fig.~2 are reported without accidental-coincidence subtraction, reflecting the raw source performance. Accidental subtraction was applied in the state-analysis measurements, the polarization tomography and phase-sensitive fringe data.
\\
Figure~\ref{fig:singles_cc} shows the measured coincidence and singles rates as functions of pump power. Accidental coincidences, enhanced by the photoluminescence background, contribute to the measured coincidence rate. After subtracting the estimated accidental contribution, we perform a linear fit constrained to pass through the origin, obtaining a detected brightness of $(10.2\pm0.4)\,\mathrm{s}^{-1}\mathrm{mW}^{-1}$. Without accidental subtraction, the corresponding slope is $(11.3\pm0.6)\,\mathrm{s}^{-1}\mathrm{mW}^{-1}$. Normalizing the corrected brightness to the nominal 40-nm detection bandwidth gives $0.254\pm0.009\,\mathrm{s}^{-1}\mathrm{mW}^{-1}\mathrm{nm}^{-1}$. 

\subsection{Spatial scan}

A spatial scan of the single-photon counts across the sample confirms that the optical signal originates from the targeted flake. The flake appears as a region of modified transmission under 426~nm excitation, while the surrounding substrate remains comparatively nonlinear weak in the detected near-infrared band. Residual background counts away from the flake are attributed to weak broadband luminescence and detector background. All spatial scans were acquired using narrowband detection centered near the SPDC wavelength, with either an $850\pm20$~nm band-pass filter or an $852\pm1.5$~nm band-pass filter. The notation $850\pm20$~nm means that the filter is centered at 850~nm and transmits approximately a 40~nm-wide spectral window, while $852\pm1.5$~nm denotes a much narrower 3~nm-wide window.

\section{Polarization-resolved tomography}
\label{sec:tomo}

\subsection{State space and density matrix}
The reconstructed state is a two-photon polarization state. A density matrix is the object that contains the complete polarization  state information accessible to the measurement. In the simplest language, its diagonal elements tell us how much population is present in each basis state, while its off-diagonal elements tell us whether different basis states are phase coherent. This is why the density matrix is the natural representation here: it can describe both an ideal pure state and a realistic experimental state that contains background, imperfect coherence, or residual distinguishability we target to probe in main text \cite{James2001,NielsenChuang2010,Lvovsky2018}. A simple two-qubit description is not sufficient if one wants to retain sensitivity to partial distinguishability and hidden degrees of freedom. We therefore use a multiparticle polarization tomography framework in which two-photon polarization states are described in a basis that separates the symmetric subspace from the antisymmetric singlet contribution \cite{Adamson2007Hidden}.
For two photons, for the convenience we use basis:
\begin{equation}
\left\{
\ket{HH},
\ket{\psi^+},
\ket{VV},
\ket{\psi^-}
\right\},
\end{equation}
where $H$ and $V$ denote horizontal and vertical polarization, following the standard single-photon polarization basis \cite{ChekhovaBanzer2021,Lvovsky2018}. The state $\ket{HH}$ means that both photons are horizontally polarized, while $\ket{VV}$ means that both photons are vertically polarized. The two mixed-polarization states are
\begin{align}
\ket{\psi^+} &= \frac{1}{\sqrt{2}}\left(\ket{HV}+\ket{VH}\right),\\
\ket{\psi^-} &= \frac{1}{\sqrt{2}}\left(\ket{HV}-\ket{VH}\right).
\end{align}
Here $\ket{HV}$ denotes one photon in $H$ and one photon in $V$, and $\ket{VH}$ denotes the same polarization content with the photon labels exchanged. The plus sign in $\ket{\psi^+}$ gives the symmetric combination, whereas the minus sign in $\ket{\psi^-}$ gives the antisymmetric combination.

The first three states span the symmetric spin-1 subspace, while $\ket{\psi^-}$ is the antisymmetric singlet state. The source-generated states are expected to be predominantly confined to the symmetric subspace. However, retaining the full $4\times4$ reconstruction is important because hidden distinguishability can populate, or effectively couple to, the antisymmetric sector.

In this basis, the reconstructed density matrix has the general form
\begin{equation}
\rho =
\begin{pmatrix}
\rho_{HH,HH} & \rho_{HH,\psi^+} & \rho_{HH,VV} & \rho_{HH,\psi^-} \\
\rho_{\psi^+,HH} & \rho_{\psi^+,\psi^+} & \rho_{\psi^+,VV} & \rho_{\psi^+,\psi^-} \\
\rho_{VV,HH} & \rho_{VV,\psi^+} & \rho_{VV,VV} & \rho_{VV,\psi^-} \\
\rho_{\psi^-,HH} & \rho_{\psi^-,\psi^+} & \rho_{\psi^-,VV} & \rho_{\psi^-,\psi^-}
\end{pmatrix}.
\end{equation}
The symbol $\rho$ denotes the density matrix. Each matrix element carries two labels: the first label identifies the row basis state, and the second label identifies the column basis state. For example, $\rho_{HH,VV}$ is the coherence between the $\ket{HH}$ and $\ket{VV}$ components. Diagonal elements such as $\rho_{HH,HH}$ are real and represent populations. Off-diagonal elements can be complex, and their phases encode relative phases between quantum amplitudes.

\subsection{Measurement operators and count normalization}

Each tomographic setting $m$ is defined by the quarter and half-wave plate angles $\theta_Q$ and $\theta_H$ listed in Table~\ref{tab:angles}., which implement a polarization transformation $U_m=U(\theta_Q,\theta_H)$. For two photons, the corresponding transformation is $U_m\otimes U_m$. The measurement operators are therefore
\begin{equation}
\Pi_{\mu,m} =
(U_m\otimes U_m)^\dagger
\Pi_\mu^{(0)}
(U_m\otimes U_m),
\end{equation}
where $\mu\in\{HH,HV_\Sigma,VV\}$ labels the polarization outcomes. The corresponding probabilities are given by $p_{\mu,m}=\mathrm{Tr}(\rho\Pi_{\mu,m})$. The combined $HV_\Sigma$ outcome includes both symmetric and antisymmetric polarization contributions.
Before reconstructing the density matrix, we subtract accidental coincidences and correct for relative detection efficiencies. The relative efficiency factors are estimated from the mean singles rates referenced to the one of the four channels.

In our detection geometry, channels 1 and 2 correspond to the H-polarized output (transmission), while channels 4 and 5 correspond to the V-polarized output (reflective port). The corrected coincidence counts are combined as
\begin{align}
N_{HH} &= 2\widetilde{C}_{12},\\
N_{HV_\Sigma} &= \widetilde{C}_{14}+\widetilde{C}_{15}
+\widetilde{C}_{24}+\widetilde{C}_{25},\\
N_{VV} &= 2\widetilde{C}_{45},
\end{align}
where $\widetilde{C}_{ij}$ denotes the corrected coincidence counts between channels $i$ and $j$. The factors of two account for the 50:50 splitting of two photons entering the same PBS output. For each analyzer setting, the measured probabilities are obtained by normalizing these three outcomes to their sum.

\subsection{Density-matrix reconstruction}

Following the multiparticle polarization tomography formalism of Adamson \textit{et al.}~\cite{Adamson2007Hidden}, we describe the symmetric polarization manifold in the
basis $\{\ket{HH},\ket{\Psi^+},\ket{VV}\}$. The normalized $3\times3$ density matrix $\rho_{\mathrm{sym}}$ is obtained by minimizing
\begin{equation}
\chi^2 =
\sum_{m,\mu}
\left[
p_{\mu,m}^{(\mathrm{meas})}
-\mathrm{Tr}(\rho_{\mathrm{sym}}
\Pi_{\mu,m}^{(\mathrm{sym})})
\right]^2,
\end{equation}
where $\Pi_{\mu,m}^{(\mathrm{sym})}$ denotes the measurement operator restricted to the symmetric sector.
The fit is constrained to produce a  positive  density matrix with unit trace.

The fidelity, purity, and spin-1 Wigner functions reported in the main text are calculated from the normalized symmetric density matrix.
The antisymmetric population is extracted from the reconstructed $4\times4$ density matrix in the basis $\{\ket{HH},\ket{\Psi^+},\ket{VV},\ket{\Psi^-}\}$ as $P_{\Psi^-}=\rho_{44}=\bra{\Psi^-}\rho\ket{\Psi^-}$.

Statistical uncertainties are estimated using  Monte Carlo resampling of the measured coincidence
counts according to Poisson statistics. The density matrix is reconstructed for each realization, and the standard deviations of the resulting fidelity, purity, and antisymmetric population distributions are reported as uncertainties.


\section{Hamiltonian derivation of source-generated polarization-\NOON{} and Holland-Burnett states}
\label{sec:hamiltonian_derivation}

This section derives the polarization state generated by SPDC in
3R-MoS$_2$ directly from the second-order interaction SPDC Hamiltonian.  
We first include allowed polarization channels, then consider the collinear, degenerate configuration where both photons are collected in same optical mode. Derivations ends up with identification of polarization basis that is selected by the laser pump. 

The central two-photon result  shown below is:
\begin{equation}
\boxed{
\ket{\Psi_2(\theta)}
=
\frac{1}{\sqrt{2}}
\left(
\ket{2,0}_{b_+(\theta),b_-(\theta)}
-
\ket{0,2}_{b_+(\theta),b_-(\theta)}
\right),
}
\label{eq:si_final_noon_state_intro}
\end{equation}
where the orthogonal polarization modes $b_\pm(\theta)$ are selected by the pump polarization.  In the complementary modes $c_\pm(\theta)=[b_+(\theta)\pm b_-(\theta)]/\sqrt{2}$, the same state is
the twin-Fock state $\ket{1,1}_{c_+(\theta),c_-(\theta)}$.  Thus, the two-photon component is simultaneously
a polarization-\NOON{} state in the $b_\pm$ basis and a twin-Fock state in the complementary
$c_\pm$ basis.

\subsection{Coordinates and tensor convention}

We choose the in-plane material axes x and y along the armchair (AC) and y along the zigzag (ZZ) direction, as indicated in Fig 1 in the main text, crystallographic $c$-axis is normal to the MoS$_2$ layers.
\begin{equation}
x\equiv \mathrm{AC},
\qquad
y\equiv \mathrm{ZZ},
\qquad
z\parallel c,
\end{equation}
The pump and the collected down-converted polarization modes are taken to lie in the layer plane, so only indices $x,y$ are retained below.

The linear pump polarization at the sample is
\begin{equation}
\hat{\mathbf e}_p(\theta)
=
\cos\theta\,\hat{\mathbf x}
+
\sin\theta\,\hat{\mathbf y}.
\label{eq:si_pump_polarization}
\end{equation}
where $\theta$ is the optical polarization angle at the sample, not the mechanical angle of the pump half-wave plate. (HWP1). For the in-plane second-order susceptibility we choose
\begin{equation}
\chi^{(2)}_{xxx}=+\chi,
\qquad
\chi^{(2)}_{xyy}
=\chi^{(2)}_{yxy}
=\chi^{(2)}_{yyx}
=-\chi,
\label{eq:si_tensor_nonzero}
\end{equation}
and
\begin{equation}
\chi^{(2)}_{xxy}
=\chi^{(2)}_{xyx}
=\chi^{(2)}_{yxx}
=\chi^{(2)}_{yyy}
=0.
\label{eq:si_tensor_zero}
\end{equation}
this choice of x and y fixes the tensor convention used below. A rotation of the material frame gives an equivalent description, if we keep transformations of tensor and pump consistent with each other. 

\subsection{SPDC Hamiltonian before and after applying crystal symmetry}

After selecting targeted/relevant optical modes, the second order interaction has the general form:
\cite{gerry2023introductory,fox2006quantum,couteau2018spdc}
\begin{equation}
\hat H_I
\propto
\sum_{\mu,\nu,\lambda\in\{x,y\}}
\chi^{(2)}_{\mu\nu\lambda}
\hat a_{p,\mu}
\hat a^\dagger_{s,\nu}
\hat a^\dagger_{i,\lambda}
+\mathrm{H.c.}
\label{eq:si_mode_hamiltonian_general}
\end{equation}
The first index denotes the pump polarization, while the other two denote the signal and idler polarizations. Without further constraints, eight in plane terms are contained in the sum,.
\begin{align}
\hat H_I\propto{}&
\chi^{(2)}_{xxx}\hat a_{p,x}\hat a^\dagger_{s,x}\hat a^\dagger_{i,x}
+\chi^{(2)}_{xxy}\hat a_{p,x}\hat a^\dagger_{s,x}\hat a^\dagger_{i,y}
+\chi^{(2)}_{xyx}\hat a_{p,x}\hat a^\dagger_{s,y}\hat a^\dagger_{i,x}
+\chi^{(2)}_{xyy}\hat a_{p,x}\hat a^\dagger_{s,y}\hat a^\dagger_{i,y}
\nonumber\\
&+
\chi^{(2)}_{yxx}\hat a_{p,y}\hat a^\dagger_{s,x}\hat a^\dagger_{i,x}
+\chi^{(2)}_{yxy}\hat a_{p,y}\hat a^\dagger_{s,x}\hat a^\dagger_{i,y}
+\chi^{(2)}_{yyx}\hat a_{p,y}\hat a^\dagger_{s,y}\hat a^\dagger_{i,x}
+\chi^{(2)}_{yyy}\hat a_{p,y}\hat a^\dagger_{s,y}\hat a^\dagger_{i,y}
+\mathrm{H.c.}
\label{eq:si_all_eight_terms}
\end{align}
For the 3R-MoS2, as stated in plane tensor satisfies relation below:
\begin{equation}
\chi^{(2)}_{xxx}
=-\chi^{(2)}_{xyy}
=-\chi^{(2)}_{yxy}
=-\chi^{(2)}_{yyx}\equiv\chi,
\label{eq:si_tensor_nonzero}
\end{equation}
So keeping all non-zero contributions \eqref{eq:si_tensor_nonzero} become:

\begin{align}
\hat H_I\propto\chi\Big(&
\hat a_{p,x}\hat a^\dagger_{s,x}\hat a^\dagger_{i,x}
-\hat a_{p,x}\hat a^\dagger_{s,y}\hat a^\dagger_{i,y}
\nonumber\\
&-
\hat a_{p,y}\hat a^\dagger_{s,x}\hat a^\dagger_{i,y}
-\hat a_{p,y}\hat a^\dagger_{s,y}\hat a^\dagger_{i,x}
\Big)+\mathrm{H.c.}
\label{eq:si_tensor_hamiltonian}
\end{align}
The selection rule is therefore already visible without choosing an
output basis:
\begin{equation}
x\text{-polarized pump}:\;xx-yy,
\qquad
y\text{-polarized pump}:\;-(xy+yx).
\label{eq:si_channels_by_pump}
\end{equation}
It is clear that x polarized pump will drive xx-yy, while a y-polarized pump drives -(xy+yx). These contributions are fixed by crystal symmetric, and the laser pump orientation will determine their relative weights.

In the low-gain/undepleted pump approximation, the coherent pump field is treated classically, we use:
\begin{equation}
\hat a_{p,x}\rightarrow \alpha_p\cos\theta,
\qquad
\hat a_{p,y}\rightarrow \alpha_p\sin\theta.
\label{eq:si_classical_pump}
\end{equation}
Considering the absorbed pump amplitude and overlap of optical mode into a generally complex coupling $\eta$ gives:
\begin{align}
\hat H_I^{(+)}(\theta)=\eta\bigl\{&
\cos\theta\,[\hat a^\dagger_{s,x}\hat a^\dagger_{i,x}
-\hat a^\dagger_{s,y}\hat a^\dagger_{i,y}]
\nonumber\\
&-\sin\theta\,[\hat a^\dagger_{s,x}\hat a^\dagger_{i,y}
+\hat a^\dagger_{s,y}\hat a^\dagger_{i,x}]\bigr\},
\label{eq:si_labeled_hamiltonian}
\end{align}

For distinguishable signal and idler mode labels, the creation part of
Eq.~\eqref{eq:si_labeled_hamiltonian} acting on vacuum produces the normalized state, where $\hat H_I^{(+)}$ denotes the pair-creation part of $\hat H_I$.
\begin{align}
\ket{\Psi_{s,i}(\theta)}
=\frac{1}{\sqrt{2}}\Big[&
\cos\theta
\left(
\ket{x_sx_i}-\ket{y_sy_i}
\right)
\nonumber\\
&-
\sin\theta
\left(
\ket{x_sy_i}+\ket{y_sx_i}
\right)
\Big].
\label{eq:si_labeled_state_normalized}
\end{align}
The four labelled kets are orthonormal, and the unnormalized squared norm
is $2\cos^2\theta+2\sin^2\theta=2$.  Equation
\eqref{eq:si_labeled_state_normalized} is the natural Bell-state
description The signal and idler labels here still identify distinct subsystems. Our experiment instead collects the degenerate photons in one spatial mode.

For the single mode description we identify the signal and idler creation operator 
\begin{equation}
\hat a^\dagger_{s,x}=\hat a^\dagger_{i,x}\equiv\hat a_x^\dagger,
\qquad
\hat a^\dagger_{s,y}=\hat a^\dagger_{i,y}\equiv\hat a_y^\dagger.
\label{eq:si_remove_labels}
\end{equation}

Bosonic commutation gives
\begin{equation}
\hat a_x^\dagger\hat a_y^\dagger
=
\hat a_y^\dagger\hat a_x^\dagger,
\end{equation}
so the $xy$ and $yx$ alternatives become the same Fock amplitude and add
coherently.  So we can define also the the pair-creation operator
\begin{equation}
\boxed{
\hat G^\dagger(\theta)
=
\cos\theta
\left[
(\hat a_x^\dagger)^2-(\hat a_y^\dagger)^2
\right]
-2\sin\theta\,\hat a_x^\dagger\hat a_y^\dagger .
}
\label{eq:si_pair_creation_operator}
\end{equation}
That gives:
\begin{equation}
\hat H_I(\theta)=\eta\hat G^\dagger(\theta)
+\eta^*\hat G(\theta).
\label{eq:si_fixed_basis_hamiltonian}
\end{equation}
The factor of two comes from adding the indistinguishable xy and yx amplitudes. 

In the weak-gain limit,
\begin{equation}
\ket{\psi_{\mathrm{out}}}
\simeq
\ket{0}
-\frac{\mathrm{i}}{\hbar}
\int\!\mathrm{d}t\,\hat H_I(t)\ket{0}
+O(|\eta|^2).
\label{eq:si_low_gain_expansion}
\end{equation}
The vacuum amplitude determines the pair-production probability, whereas
the normalized two-photon component determines the conditional
polarization state.  Putting in:
\begin{equation}
(\hat a_x^\dagger)^2\ket{0}=\sqrt{2}\ket{2,0}_{x,y},
\quad
(\hat a_y^\dagger)^2\ket{0}=\sqrt{2}\ket{0,2}_{x,y},
\quad
\hat a_x^\dagger\hat a_y^\dagger\ket{0}=\ket{1,1}_{x,y},
\label{eq:si_fock_normalization}
\end{equation}
we obtain the pair-creation operator:
\begin{equation}
\hat G^\dagger(\theta)\ket{0}
=2\left[
\frac{\cos\theta}{\sqrt{2}}
\left(
\ket{2,0}_{x,y}-\ket{0,2}_{x,y}
\right)
-\sin\theta\ket{1,1}_{x,y}
\right].
\label{eq:si_fixed_basis_unnormalized}
\end{equation}
which results in a normalized two photon component. 
\begin{equation}
\boxed{
\ket{\Psi_2(\theta)}
=
\frac{\cos\theta}{\sqrt{2}}
\left(
\ket{2,0}_{x,y}-\ket{0,2}_{x,y}
\right)
-
\sin\theta\ket{1,1}_{x,y}.
}
\label{eq:si_fixed_basis_state}
\end{equation}
The expression above describes the state varying polarization we observe, in a fixed material basis as the pump is rotated. 

\subsection{Pump selected N00N basis}

Equation~\eqref{eq:si_pair_creation_operator} can be
written as the quadratic form
\begin{equation}
\hat G^\dagger(\theta)
=
\begin{pmatrix}
\hat a_x^\dagger & \hat a_y^\dagger
\end{pmatrix}
\underbrace{
\begin{pmatrix}
\cos\theta & -\sin\theta\\
-\sin\theta & -\cos\theta
\end{pmatrix}}_{M(\theta)}
\begin{pmatrix}
\hat a_x^\dagger\\
\hat a_y^\dagger
\end{pmatrix}.
\label{eq:si_quadratic_matrix}
\end{equation}

Since $M(\theta)^2=\mathbb{I}$, its eigenvalues are $+1$ and $-1$. The corresponding orthogonal
polarization modes are
\begin{equation}
\boxed{
\begin{aligned}
\hat a^\dagger_{b_+(\theta)}
&=
\cos\frac{\theta}{2}\,\hat a_x^\dagger
-
\sin\frac{\theta}{2}\,\hat a_y^\dagger,
\\
\hat a^\dagger_{b_-(\theta)}
&=
\sin\frac{\theta}{2}\,\hat a_x^\dagger
+
\cos\frac{\theta}{2}\,\hat a_y^\dagger.
\end{aligned}
}
\label{eq:si_b_basis_creation}
\end{equation}

This transformation can be verified by expanding the creation operator quadratically:
\begin{align}
&(\hat a^\dagger_{b_+(\theta)})^2
-(\hat a^\dagger_{b_-(\theta)})^2
\nonumber\\
&\quad=
\cos\theta
\left[
(\hat a_x^\dagger)^2-(\hat a_y^\dagger)^2
\right]
-2\sin\theta\,
\hat a_x^\dagger\hat a_y^\dagger
=\hat G^\dagger(\theta).
\label{eq:si_half_angle_identity}
\end{align}
The interaction Hamiltonian therefore becomes
\begin{equation}
\boxed{
\hat H_I(\theta)
=
\eta\left[
(\hat a^\dagger_{b_+(\theta)})^2
-
(\hat a^\dagger_{b_-(\theta)})^2
\right]
+\mathrm{H.c.}
}
\label{eq:si_diagonal_hamiltonian}
\end{equation}

The two terms describe coherent pair creation amplitudes in the orthogonal modes $b_\pm(\theta)$, with a relative minus sign fixed by the nonlinear interaction. Applying the pair creation Hamiltonian to the vacuum and normalizing gives
\begin{equation}
\boxed{
\ket{\Psi_2(\theta)}
=
\frac{1}{\sqrt{2}}
\left(
\ket{2,0}_{b_+(\theta),b_-(\theta)}
-
\ket{0,2}_{b_+(\theta),b_-(\theta)}
\right).
}
\label{eq:si_source_generated_noon}
\end{equation}
SO, for any linearly polarized pump, the generated biphoton is an $N=2$ polarization-N00N state in the pump-selected basis. Changing the pump polarization rotates this basis while preserving the  necessary N00N state structure.

\subsection{Complementary \texorpdfstring{$c_\pm$}{c modes}: twin-Fock and higher number pairs}

We now define another complementary polarization modes
\begin{equation}
\boxed{
\hat a^\dagger_{c_+(\theta)}
=
\frac{
\hat a^\dagger_{b_+(\theta)}+
\hat a^\dagger_{b_-(\theta)}}{\sqrt{2}},
\qquad
\hat a^\dagger_{c_-(\theta)}
=
\frac{
\hat a^\dagger_{b_+(\theta)}-
\hat a^\dagger_{b_-(\theta)}}{\sqrt{2}}.
}
\label{eq:si_c_basis_creation}
\end{equation}
The modes $c_\pm(\theta)$ form the complementary twin-Fock basis shown
in Fig.~1.  They should not be confused with the laboratory H/V labels. They are rotated by $45^\circ$ relative to $b_\pm(\theta)$ as can be seen in Fig. 1(a) left.
\begin{equation}
(\hat a^\dagger_{b_+(\theta)})^2
-(\hat a^\dagger_{b_-(\theta)})^2
=
2\hat a^\dagger_{c_+(\theta)}
\hat a^\dagger_{c_-(\theta)}.
\label{eq:si_difference_squares_c}
\end{equation}
and we also have:
\begin{equation}
\boxed{
\ket{\Psi_2(\theta)}
=
\ket{1,1}_{c_+(\theta),c_-(\theta)}.
}
\label{eq:si_noon_as_twin_fock}
\end{equation}
The n00n and twin-Fock expressions are the same state written in two complementary polarization basis. IN the usual Holland-Burnett construction, a twin Fock input is mixed on balanced beam splitter. The resulting state is the Holland-Burnett state. 

In the standard Holland-Burnett construction, a twin-Fock state $\ket{n,n}$ is incident on a balanced beam splitter, and the state in the two output modes is called a Holland-Burnett state \cite{HollandBurnett1993,Datta2011HB}. Where we emphasize here that no post source beam splitter is needed to prepare the two photon state.

The same interaction also determines the higher order photon number components. In the complementary $c_\pm(\theta)$ basis, the Hamiltonian becomes
\begin{equation}
\hat H_I(\theta)
=
2\eta\,
\hat a^\dagger_{c_+(\theta)}
\hat a^\dagger_{c_-(\theta)}
+\mathrm{H.c.},
\label{eq:si_two_mode_squeezing_c}
\end{equation}
which describes the creation of photon pairs in two orthogonal polarization modes and has the standard form of a two-mode squeezing Hamiltonian. Considering higher-order pair generation, the output from vacuum in the ideal single-mode approximation is given by~\cite{CerfJabbour2020,Birrittella2023}.

\begin{equation}
\ket{\psi_{\mathrm{out}}}
= \frac{1}{\cosh r}\sum_{n=0}^{\infty}\left(e^{i\phi}\tanh r\right)^n\ket{n,n}_{c_+(\theta),c_-(\theta)},
\end{equation}

where $r$ and $\phi$ describe the effective interaction strength and phase~\cite{CerfJabbour2020}. Each fixed $2n$ photon component is a twin Fock state in the $c_\pm$ basis, corresponding to a Holland-Burnett state in the complementary $b_\pm$ basis~\cite{HollandBurnett1993,Datta2011HB}. For $n=1$, this reduces to the two-photon N00N state derived above. For higher photon numbers, Holland-Burnett states also contain intermediate even-photon occupations and are distinct from N00N states. Such as seen below in \ref{eq:si_hb4}, where or $n=2$\cite{Datta2011HB}, has a a form of:

\begin{equation}
\begin{aligned}
\ket{\mathrm{HB}_4}
&= \frac{\sqrt{6}}{4}
\left(\ket{4,0}+\ket{0,4}\right)_{b_+,b_-} \\
&\quad - \frac{1}{2}\ket{2,2}_{b_+,b_-}.
\end{aligned}
\label{eq:si_hb4}
\end{equation}

In our 3R-MoS2 case, for any pump orientation, the generated two-photon state is a N00N state in the pump-dependent $b_\pm(\theta)$ basis and, equivalently, a twin-Fock state in the complementary $c_\pm(\theta)$ basis. For the crystallographic AC ($\theta=0$) direction, the state is an H/V N00N state, equivalently a D/A twin-Fock state, whereas for the ZZ ($\theta=\pi/2$) direction, it is an H/V twin-Fock state, equivalently a D/A N00N state.


\subsection{State labels and entanglement partition}

The Bell, \NOON{}, and twin-Fock/Holland--Burnett descriptions emphasize
different mode partitions of the same two-photon amplitude.  With
distinguishable signal and idler labels, Eq.~\eqref{eq:si_labeled_state_normalized}
is a two-qubit Bell-type state.  In the single collected degenerate mode,
the occupation-number description is natural: the state is mode
entangled in the $b_+|b_-$ partition and has the \NOON{} form
\eqref{eq:si_source_generated_noon}, while it is the definite occupation
$\ket{1,1}$ in the complementary $c_+|c_-$ partition.  This is not a
contradiction; mode entanglement is defined relative to the chosen mode
decomposition.  The basis-independent experimental statement is the
coherent two-photon polarization state reconstructed by tomography.

\section{Two-photon Wigner representation on the Poincar\'e sphere}
\label{sec:wigner}

The reconstructed two-photon polarization states are visualized using the spin-$J$ Wigner representation on the Poincar\'e sphere, with $J=1$ for the symmetric
two-photon manifold. Unlike a representation based only on the mean Stokes vector, the Wigner function captures higher-order polarization correlations and coherence, providing a useful visualization of the two-photon N00N states~\cite{Leonhardt1997}.

The Wigner function is expanded in spherical harmonics as
\begin{equation} 
W(\vartheta,\varphi)=\sum_{K=0}^{2J}\sum_{q=-K}^{K}\mathcal{T}_{Kq}Y_{Kq}(\vartheta,\varphi), \end{equation}
where $Y_{Kq}$ are spherical harmonics and
$\mathcal{T}_{Kq}$ are the corresponding multipole coefficients obtained from the density matrix
~\cite{Varshalovich1988}.

The experimental Wigner functions shown in the main text are calculated from the reconstructed symmetric density matrices $\rho_{\mathrm{sym}}$, while the theoretical distributions are obtained from the corresponding target states in Eq.~\eqref{eq:si_source_generated_noon}. The $H/V$ polarizations define the poles of the Poincar\'e sphere, while $D/A$ lie on its equator. The angular structure of the Wigner distributions visualizes the two-photon coherence and its evolution with pump polarization, providing a complementary representation of the reconstructed density matrices.


\newpage

\bibliographystyle{apsrev4-2}
\bibliography{references}